# Reorganization of Links to Improve User Navigation

Deepshree A. Vadeyar[1], Yogish H.K[2]

[1]Department of Computer Science and Engineering, EWIT Bangalore
deepshri.b@gmail.com

[1]Department of Computer Science and Engineering, EWIT Bangalore
yogishhk@gmail.com

## ABSTRACT

*Website can be easily design but to efficient user navigation is not a easy task since user behavior is keep changing and developer view is quite different from what user wants, so to improve navigation one way is reorganization of website structure. For reorganization here proposed strategy is farthest first traversal clustering algorithm perform clustering on two numeric parameters and for finding frequent traversal path of user Apriori algorithm is used. Our aim is to perform reorganization with fewer changes in website structure.*

## 1. INTRODUCTION

WWW is large source of information and lakhs of user uses internet as search engine and website, website is used for providing information and also it is big source of commercialization but even user not get proper information over website and seeking for page what user wants, its happening due to reason that user view to use website is different than developer and as user has different characteristics than other user [1].As user have different requirement so how to naviage use effectively one of the way is by changing link structure[2] ,by changing structure we can say that we are performing web transformation .Web transformation is not done manually but ats automatic process by providing some learning method to websites, we used un-supervised learning method Clustering and intelligent method Association to find frequent links by traversing path of user.

As we performing data mining techniques over website we can define it as web mining which categorized as web usage mining in this technique weblogs are collected according to user behavior, web content mining is about content on web pages and it is also called as Text mining and web structure mining about changing link structure. As we are performing link mining but it depends on the web usage data and we also performed some of pre-process task.

Our work is to perform clustering for this we chooses two parameter Average duration of user we can say time for which user on website and number of clicks on URL of web pages, we perform clustering for the parameter more than some defined threshold ,as more click on URL we can get cluster with high frequency of page access, due to clustering large data space say 6000 rows divide is some clusters and each cluster may conation 100,1000 rows according to similarity measures, as we used here farthest first traversal clustering algorithm our clustering performed on Euclid and Manhattan similarity measures and objects are assign at smallest distance of objects from centroid and max distance between centroid actually centrod is chosen as randomly bust with maximum Euclid distance between two centroid and distance between centroid to object is minimum distance.

To find which link structure to be change we will user identification and we also collect traversing path of user for which after performing association by using Apriori algorithm, we will get frequent links, if frequent links are available in cluster means user are on page for more time and we have pages with high frequency, we will check it for categorization and with out-degree threshold and if both condition satisfies link will be reorganized.

Categorization is nothing but storing link structure as binary 0 or 1 as absence and presence of structure respectively so we can check whether link already present, we will use this as matrix so out-degree can be calculates easily.

As we used Apriori algorithm with threshold min-support, max-support, min-confidence and delta later we will get best item sets say x-item set 'L(X)',and best rule found which can be used to find best link for which reorganization can be performed in later section we will see how the item sets and formed and that search on cluster to perform link mining., at last web pages should be reorganized in such way that it should full fill the user needs

In later section we shown how links will be structured and we also performed comparison of other clustering algorithm. We found that farthest first is fastest in building model. We can also conclude that when links which to be reorganized available in clusters and frequent item set as data is reduced before clustering by data reduction and outlier mining then time to reorganized also reduced.For website navigation as our first step is on weblogs so before actual link mining we need to perform some pre-processing and for this we considered some threshold values like session-threshold 'α' and click-threshold 'β' here α defines time spent more than threshold considered in clustering and β define click on links more than threshold consider for clustering. Following are steps for pre-processing :

(1)Data cleaning: Removal of session value and clicks less than define α and β.

(2)User Identification: Finding user with unique ip even user on same network so if we get frequent user we can provide priority to user .

(3) Session Identification: Method of finding average time of user spent on website.

(4)Path Completion: Finding missing links but which are important and frequently request.

(5) Formatting: Converting logs into data which can be mined. It includes removal of binary value for clustering and removal of numeric parameter for frequent mining.

This pre-processing task actually removes irrelevant links before start of miming so time of building model and reorganizing of links will be reduced.

As our aim is to improve user navigation so we proposed term Improved efficiency that find how much user navigation is improved after reorganization, improved efficiency can be find by below equation.

$$\mathrm{Im}\, proved\, Eficiency = \frac{T_p - P_t}{T_p} * 100 \quad \text{----------(1)}$$

In equation '$T_p$' is total path taken by user and '$P_t$' is path taken by user on each link in this way equation give navigation efficiency after reorganization. For example if we consider that some links are directly connected to initial page than its efficiency is improved by 1 path then here value of path taken i.e. $P_t=1$ so efficiency equation can be modify as given below.

$$\text{Im } proved\ Efficiency = \frac{T_p - 1}{T_p} * 100 \quad \text{--------(2)}$$

## 2. RELATED WORK.

In this section we are providing information about previous work done for web site intelligence in terms of web usage mining and web transformation, our main focus on link mining for which we first need input parameters as web logs and data mining technique over web logs called as web usage mining. Introduction of web usage mining by Cooley in [3]explains about usage mining techniques, for working with usage mining author proposed several mining techniques, their main focus is on some data preparation and pre-processing techniques for web logs .

Perkowitz in [4] focus on problem of page indexing and for this proposed technique is clustering like novel and conceptual clustering algorithm(COBWEB).In this paper clustering of objects are achieved by means of some common content shared between objects, here author also proposed quality measures like user looking for page and efforts by user to get desired page later they used evaluation by measures like how much website improved and how many users are benefitted.

As our main goal is structure mining and many work done on structure mining as Joy Shalom in [5] consider website structure as graph and website reorganization can be achieved by means of server logs ,proxy server and cookies, other main criteria suggested by author is browsing efficiency that is ratio of shortest path from index page to desired page to cost of operation. Here main aim is to guide user for accessing website effectively.

Lin in [6] proposed here a model for reducing load on searching desired page for user. Their approach is based on session and pages which occurred together like if 'A' access than 'B' access than sequence from A to B is important here.

Fu in [7]proposed reorganization of website based on used access pattern and here to achieve change in structure strategies used are pre-processing and classification. Here in pre-processing step information about website and web server logs is maintained. Second step include page classification based on content of website here we can say content mining is involved on parameters like file type, total links present ,session present and user time on website.

Gupta in[8] proposed reorganization by classification techniques based on type of file extension, number of links page, ratio of session on last page to the total session on web site and average time for which user on websites or user is login.

Min chen in [9] proposed model for reorganizing website structure for reorganizing website structure with minimum change ,here to know how much changes required out-degree threshold is used, their model also consist of path threshold,their model work on traversal path required to reach target page.

Yitong in [10] proposed k-means clustering algorithm using cosine similarities for link analysis and objects are cluster according to common links, words or phrases shared between documents. They used here concept of co-citation and coupling with concept of Hubs and Authority.

Amar Singh In [11] proposed clustering approach over links for improving searching of links in search engines and to accomplish this they used k-means and page-rank algorithm. Later they also shown results for weighted page rank algorithm using k-means.

Mobasher In[12][13][14] proposed clustering methods based on usage mining. In [12] author first find frequent item set and then performed clustering on user profiles. In[13]author reverse process first cluster and then finf frequent item set on usage data so imp data not lost from analysis. In [14] author proposed clustering algorithm on parameter like user transaction and page views.

## 3. OUR MODEL

We consider websites as graph and each page as node and redirecting URL between pages as edges figure-1 shows our website with 15 pages and many links, links as edges represent as 1 or 0 say we have source node is i and node on which out-link connect to j node with link $X_{ij}$ so this can be represent as equation 3.

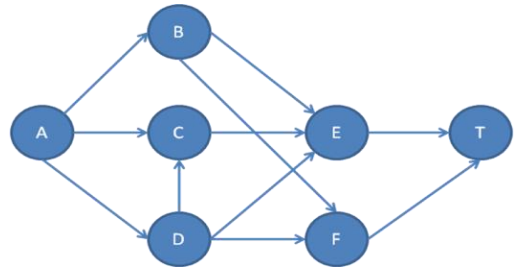

Figure-1: Website structure

$$X_{ij} = 1 \quad \text{Or} \quad X_{ij} = 0 \quad \text{-------(3)}$$

In equation 1 represents that there is links from page i to page j and zero represents there is no links between pages, it means user can traverse to connected links but if links is not there and page is most visited or user spent more time tan link can be created by reorganizing its website structure.

Here we will form cluster for which we have object are URL on which user is active and URL which satisfies threshold criteria for cluster of links between node i and j for which cluster can be represent as $K_{ij}$, to perform clustering we will use the similarity measures by farthest distance and we use distance measures is Euclid, here we have two parameter session 'S' and Clicks 'C' on which cluster is performed for most far value from mean as shown in equation 4.

$$d((i,j)) = \sqrt{(S_i - S_j)^2 + (C_i - C_j)^2} \quad \text{-------(4)}$$

We also considered the user behaviour and time for which user spent more time on pages we consider that page as target page and we saves user navigation path. Say 'I' is item set contain 'n' item set like $I = I_1, I_2, ---- I_n$ and each item shows navigation path of user which can be $I_1 = (A,B,E,K), I_2 = (A,C,J,K), I_n = (A,B,E,A,J,K)$ here we can show that 'K' is more frequent and user traversing back from 'E' to 'A' that is start page, as on user navigated path we perform Apriori algorithm, if we have user with say 'I' item set with 'N' rows after association we get frequent URL of item sets which is less than original set sat L(k)

After association we need to check the links which are available in association $A_{ij}$ as well as in clusters $C_{ij}$ so we can say there is mapping between Links 'i and j' in cluster and in Item-set that can be represent as equation 5. $A_{ij} \rightarrow C_{ij}$ .-------(5)

If links are matches between frequent item set and the clusters than we considered that links as links to be reorganized. To achieve minimum changes in website structure we considered an out-degree threshold which defines how many links can be change that can be determine by categorization. For example out degree threshold is four so only four out-links are allowed on page if already four links present we can't make out-links from that page.

## 3.1 Farthest First Algorithm

Farthest first algorithm proposed by Hochbaum and Shmoys 1985 has same procedure as k-means, this also chooses centroids and assign the objects in cluster but with max distance and initial seeds are value which is at largest distance to the mean of values, here cluster assignment is different, at initial cluster we get link with high Session Count, like at cluster-0 more than in cluster-1, and so on.
Farthest first algorithm need less adjustments and basic for this explained in [15].
 Working as described here, it also defines initial seeds and then on basis of 'k' number of cluster which we need to know prior. In farthest first it takes point $P_i$ then chooses next point $P_1$ which is at maximum distance. $p_i$ is centroid and $p_1, p_2, ........p_n$ are points or objects of dataset belongs to cluster from equation 6.

$$\min\{\max dist(p_i, p_1), \max dist(p_i, p_2)......\}  \text{------------------------------------------------- (6)}$$

Farthest first actually solves problem of k-centre and it is very efficient for large set of data. In farthest first algorithm we are not finding mean for calculating centroid, it takes centrod arbitrary and distance of one centroid from other is maximum figure-2 shows cluster assignment using farthest –first. When we performed outlier detection for our dataset we get which objects is outlier.

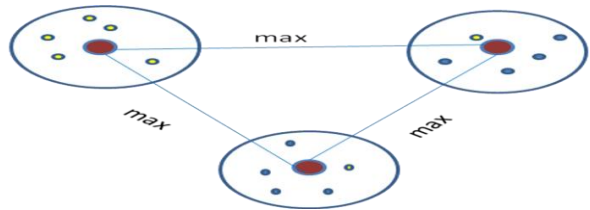

Figure-2 Object assignment in cluster

## 4. RESULT

We collected data set from Depaul university on which first we compare clustering algorithm using tool weka, fig-3 shows comparison between all algorithms and we found that farthest –first is fastest algorithm is fastest among others for building model. After clustering we performed association on transactional database, we get frequent item sets, here we performed association with delta=0.05 so time of finding frequent item-set is reduced since by using delta minimum support decrease by 0.05 in each iteration.

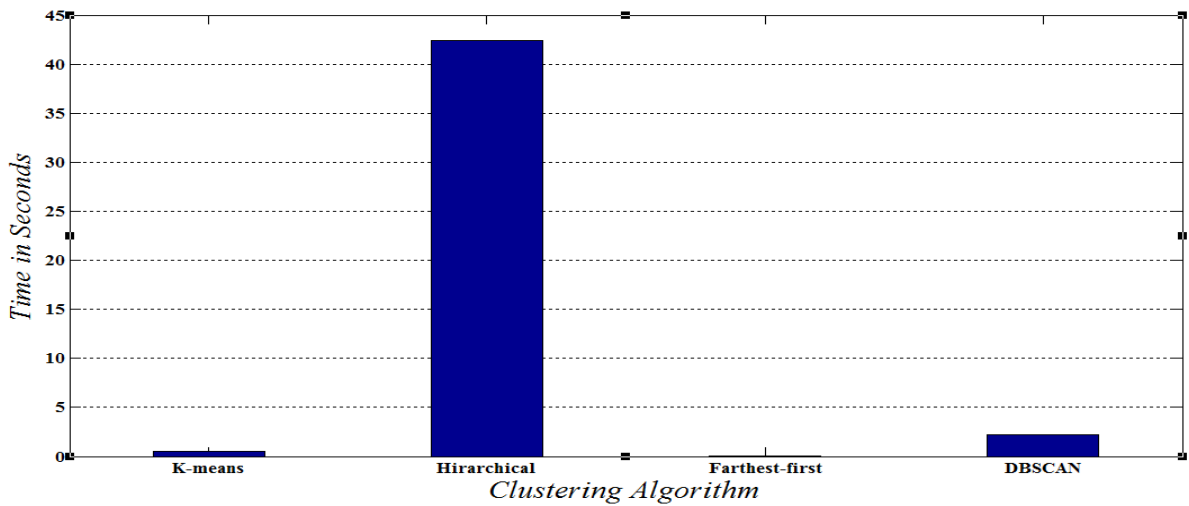

Figure-3: Comparison of clustering algorithm

Benefit of using cluster is search time for finding links to be reorganized is less than search time from transactional data base and we also get best links since links at initial clusters are links of largest duration of user on page.

Outlier analysis performed in weka tool by using filter interquartile ranges, here using farthest first outliers are more specific since this procedure not take high session_count or clicks as outliers but for k-means it takes ,since our main gal is reorganization on high count and clicks so we can say we get better outliers detection with farthest –first algorithm.

For reorganizing website structure we use real dataset on website like songs.pk , parameters session and clicks used for which links are clustered. After getting frequent item set and cluster data we found that time of reorganization using farthest first is reduced shown since time required to execute algorithm is 0.2 sec and for k-means it is 0.42 seconds, if we got assignment of links in cluster similar for farthest first and k-means even then total time of reorganization is reduced since farthest first execution time is less then k-means. Theoretical complexity for farthest first traversal algorithm is O (nk), where n is number of objects in the dataset and k is number of desired clusters. Similarly for k-means complexity is O (nkt) where n and k same as farthest first and 't' is number of iteration.

## 5. CONCLUSION

As we simulated and from result we got farthest first algorithm is fastest even than k-means and it also solves k-centre problem other aspect is better result for outlier detection we used, so it is more advantageous, hence we adopt that and when we search for links to be reorganized, we used set of traversing path for which we get matching links in cluster for which links are reorganized.

    Second usefulness of farthest first is we get links to be reorganized most near to the max parameters since it works on largest distance so we get best links. Future work remain is to find frequent item sets for particular user which we get by identification due to which mostly visiting user of website will be benefited.

    At last we can conclude as proposed algorithm gives some similar object assignment as k-means but in less time. It solves k-centre problem. If we perform evaluation then also proposed algorithm provides more correct instances for clustering.

This paper can be extended with outlier mining since proposed algorithm is faster than k-means but k-means is not robust for outliers so this future work will leads for good clustering algorithm selection

## ACKNOWLEDGEMENT

We would like to express our gratitude to all those who gave their valuable times to guide and solving doubts. We also want to thanks Computer science and Engineering department of East West Institute of Technology, Bangalore for giving me permission to continue my research on this topic.